\documentclass[]{spie}  %>>> use for US letter paper
\usepackage{amsmath,amsfonts,amssymb}
\usepackage{graphicx}
\usepackage[colorlinks=true, allcolors=blue]{hyperref}

\title{Ground control to major time-lag: on-sky results of data-driven predictive wavefront control at Keck Observatory}

\author[a]{Jules Fowler}
\author[a]{Rebecca Jensen-Clem}
\author[b]{Sylvain Cetre}
\author[c]{Maaike A. M. van Kooten}
\author[a]{Maissa Salama}
\author[d]{Antonin Bouchez}
\author[d]{Avinash Surendran}
\author[d]{Charlotte Guthery}
\author[d]{Eduardo Marin}
\author[d]{Mahawa Cisse}
\author[d]{Max Service}
\author[e]{Charlotte Z. Bond}
\author[a]{Emiel Por}
\author[f]{Nour Skaf}
\author[g]{Will Gauvin}
\affil[a]{University of California, Santa Cruz}
\affil[b]{Wakea Consulting}
\affil[c]{National Research Council Canada - Herzberg}
\affil[d]{W. M. Keck Observatory}
\affil[e]{UK Astronomy Technology Centre}
\affil[f]{University of Hawai'i}
\affil[g]{Swinburne University of Technology}

\authorinfo{Further author information: (Send correspondence to J.F.)\\J.F.: E-mail: jumfowle@ucsc.edu,}

\begin{document} 
\maketitle

\begin{abstract}
Directly imaging and characterizing exoplanets requires extreme adaptive optics (XAO), which achieves exquisite wavefront correction over a small ($<$5'') field of view. Temporal errors, where the wavefront evolves faster than the lag between wavefront sensing and control,  are often a leading term in the error budget for these XAO systems. Predictive control mitigates temporal errors by predicting where the wavefront will be by the time the system correction is applied. In particular, empirical orthogonal functions (EOF) learn linear correlations in a wavefront using previous states in the wavefront sensor history. We present on-sky results of a new implementation of EOF built directly into the Keck-II real time controller. This work revitalizes previous predictive control work at Keck, by showing improved wavefront correction from EOF as compared to a classic integral controller during on-sky observing. On-sky engineering tests at Keck Observatory of the predictive controller show a 20$\%$ performance improvement over a classic integrator according to wavefront residuals from the Shack-Hartmann Wavefront Sensor (SHWFS). Parameter optimization studies show that there is a clear improvement based on varying predictive filter hyper-parameters, but that within a reasonable regime, varying filter parameters does not degrade performance to notably worse than an integrator. 
NIRC2 imaging through the Brackett $\gamma$=2190 nm filter shows comparable performance between an integrator and predictor, both comparing Strehl Ratio (SR) and coronagraph-free contrast. We also explore power in principal components, and find a modest improvement (on the order of 3$\%$ less area under the curve of component strength) from the predictor over the integrator. This work not only improves current observing for the Keck community, but also acts as a pathfinder for predictive control methods with extremely large telescopes.
\end{abstract}

\keywords{predictive wavefront control, empirical orthogonal functions (EOF), Keck Observatory, on-sky demonstration}

\section{INTRODUCTION}
\label{sec:intro}  % \label{} allows reference to this section

Every facility adaptive optics (AO) system for an 8-10 meter class telescope uses a classic integral style controller for their high order control, which senses wavefront aberrations, estimates a correction, and applies that correction as quickly as possible. That process takes some amount of time -- typically 1-5ms, usually dominated by the time it takes a wavefront sensor to capture a high signal image of incoming turbulence. However, wind layers move faster, creating turbulence that evolves faster than the AO system can correct. This has a negative impact on coronagraphic images, which depend on high quality point spread functions (PSFs) for good corongraphic nulling. This can lead to an effect called the wind-driven halo \cite{Cantalloube2018}, where an erroneous halo is visible in the coronagraphic dark zone along a the direction of a strong wind layer. 

A natural solution is prediction; instead of applying a correction as fast as possible and hoping that is fast enough to overcome wind-layers, predictive methods estimate where the state of the atmosphere will be the correction is applied. Prediction not only combats temporal errors from fast wind layers, but also has the potential to increase signal to the wavefront sensor, by drastically reducing the error hit when running slower control loops (i.e., giving the wavefront sensor longer to expose). 

Predictive wavefront control for astronomical AO has been discussed in the literature for going on 30 years, arguably starting with Ref. \citenum{Dessenne1997}, which tested a predictor on-sky at the Observatoire de Haute-Provence. However, despite a wealth of new algorithms that range from engineering controllers (e.g., Linear Quadratic Gaussian Controllers in Ref. \citenum{Sivo2014, poyneer2007}) to multi-level neural net controllers (for example Ref. \citenum{Landman2021, Nousiainen2022}), there has yet to be a facilitized demonstration of predictive wavefront control on-sky. 

At the Subaru/SCExAO telescope, a data-driven predictive wavefront controller (Empirical Orthogonal Functions, EOF) was briefly implemented, but no work to characterize its performance was done and eventually the ability to run PWFC was lost during an RTC upgrade (private communication SCExAO team and \citenum{currie2019}). Similarly at Keck Observatory, much work went into developing a similar EOF implementation for the Keck-II AO system \cite{vanKooten2022}, which worked in concert with the Pyramid wavefront sensor as part of the KPIC instrument. This implementation showed excellent results in wavefront sensor performance and an increase in contrast by up to a factor of 3 at mid inner working angles \cite{vanKooten2022}. However, this implementation was tied to the Pyramid wavefront sensor computer, and was ultimately decommissioned from Keck-II. 

Testing predictive methods on-sky at 8-10 meter class telescopes not only improves observing quality at our current flagship observatories, but is vital to the next generation of observatories that will need this technology to reduce every potential source of error in our adaptive optics systems and find an Earth-like planet. Using large telescopes as a testbed for future observatories is the only way to demonstrate the full complexity of telescope systems. Specifically, Keck Observatory is the only segmented telescope with facility AO, making it a one-of-a-kind demonstrator for future segmented extremely large telescopes.  We discuss a new implementation of EOF built directly into the Keck-II real time controller (RTC), our efforts to commission it on the Keck-II AO bench, and final on-sky results.

\section{EMPIRICAL ORTHOGONAL FUNCTIONS AS A LINEAR DATA-DRIVEN PREDICTOR}
\label{sec:eof_def}

Empirical Orthogonal Functions (EOF), first suggested by Ref. \citenum{Guyon2017}, though closely related to the first predictive controller \cite{Dessenne1997}, uses previous states in wavefront sensor history to find linear trends in an evolving atmosphere. 

Using wavefront sensor residuals and deformable mirror commands we create pseudo-open loop (POL) records of the wavefront. If there is not error in the reconstruction, POL data represent the condition of the wavefront as if the deformable mirror correction was never applied, which linearly encodes the trends of the atmosphere. We collect n POL states in a history vector \textbf{h}, in practice 3$\leq$n$\leq$10, which will be used in real time to calculate a wavefront correction of the form: 
\begin{equation}
    \textrm{prediction}=\mathbf{F}\times\mathbf{h}
\end{equation}
where \textbf{F} is a filter matrix trained using up to a few minutes of data. 

We minimize the difference between training data \textbf{D} and a truth condition (one time lag in the future) \textbf{P} 
\begin{equation}
\left|\left|\mathbf{D}^T\mathbf{F}^T - \mathbf{P}^T\right|\right|^2
\end{equation}
and calculate \textbf{F} with a least squares inversion
\begin{equation}
\mathbf{F} = \left((\mathbf{D}^T)+ \mathbf{P}^T)^T = \mathbf{P}\mathbf{D}^T(\mathbf{D}\mathbf{D}^T + \alpha\mathbf{I}\right)^{-1}
\end{equation}
where $\alpha$ is a regularization factor, akin to setting the number of singular values. For simulations and bench experiments we are able to run with $\alpha=1$ but for on-sky we have needed higher values, which provides a more stable and less predictive correction. 

In practice, on-sky we do not run with a pure predictor, but using a mixing factor m to combine a predictor with an integrator solution, such that 
\begin{equation}
    \textrm{final correction} = g\times\left((1-m)C_i + mC_p\right)
\end{equation}
where g is the high order integrator gain, and $C_i$, $C_p$ represent the corrections calculated by the integrator and predictor respectively. 

Predictive control requires the use of multiple frames of data in real time -- in practice up to 10. Because we needed access to more frames of wavefront sensor information in real time than were available in the existing Keck real time controller (RTC) this required an RTC update. In 2024 we began discussions with Microgate and Keck to commission a new predictive wavefront control business unit in the Keck RTC. Those specifications and design documentation are described in an updated version of Ref.  \citenum{hakakaon}. For more description of this update and a control diagram see Appendix \ref{sec:bu_update}

\section{PERFORMANCE VERIFICATION WITH THE KECK-II AO BENCH}
\label{sec:bench}

After the completion of this upgrade in May of 2025, we tested the implementation on the Keck-II AO bench. During the day with the dome closed, Keck-II operates as a testbed, fed by a broadband fiber source, using the wavefront sensor, high order deformable mirror, tip/tilt mirror, and RTC that are used for nighttime on-sky operations. For the data collected on bench and on-sky, Keck was operating with a 349 degree of freedom high order Xinetics deformable mirror (DM) with each aperture mapped to one Shack-Hartmann wavefront sensor (SHWFS) subaperture, with an OCAM2k as the SHWFS detector. During natural guidestar mode (which also applies to the fiber source), tip and tilt slopes are estimated from the average tip and tilt of the SHWFS, and applied with th fast tip/tilt mirror. 
We test the predictive controller by injecting two atmospheric turbulence profiles on the DM, one with a single injected wind layer, and one with the seven layer turbulence profile based on atmospheric profiling and modeling from Ref.  \citenum{KAON303}. We inject each profile by applying the shape on the high order DM. We note that this creates turbulence that the SHWFS could perfectly sense (i.e., there are no structures at finer spatial features) and the DM could perfectly correct (i.e., there is no amplitude outside the linear range) which is not realistic to true atmospheric turbulence.

\begin{figure} [ht]
   \begin{center}
   \begin{tabular}{c} %% tabular useful for creating an array of images 
   \includegraphics[height=6cm]{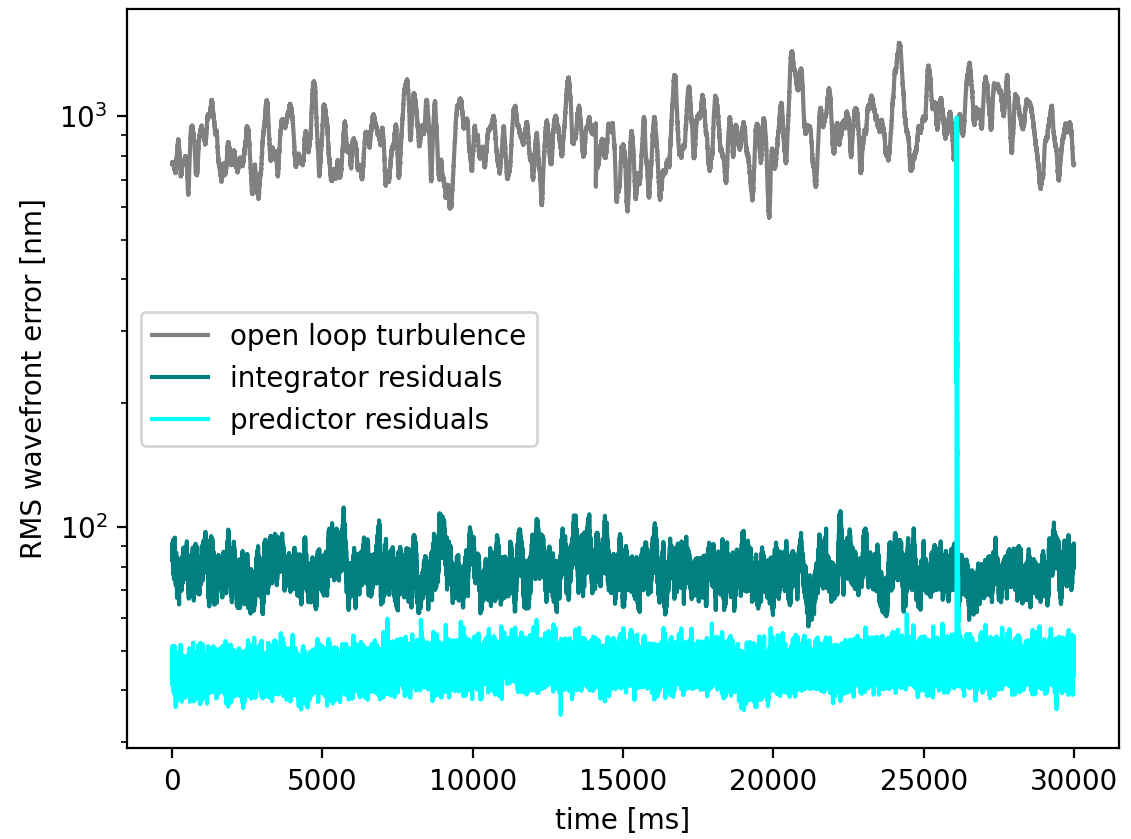}
      \includegraphics[height=6cm]{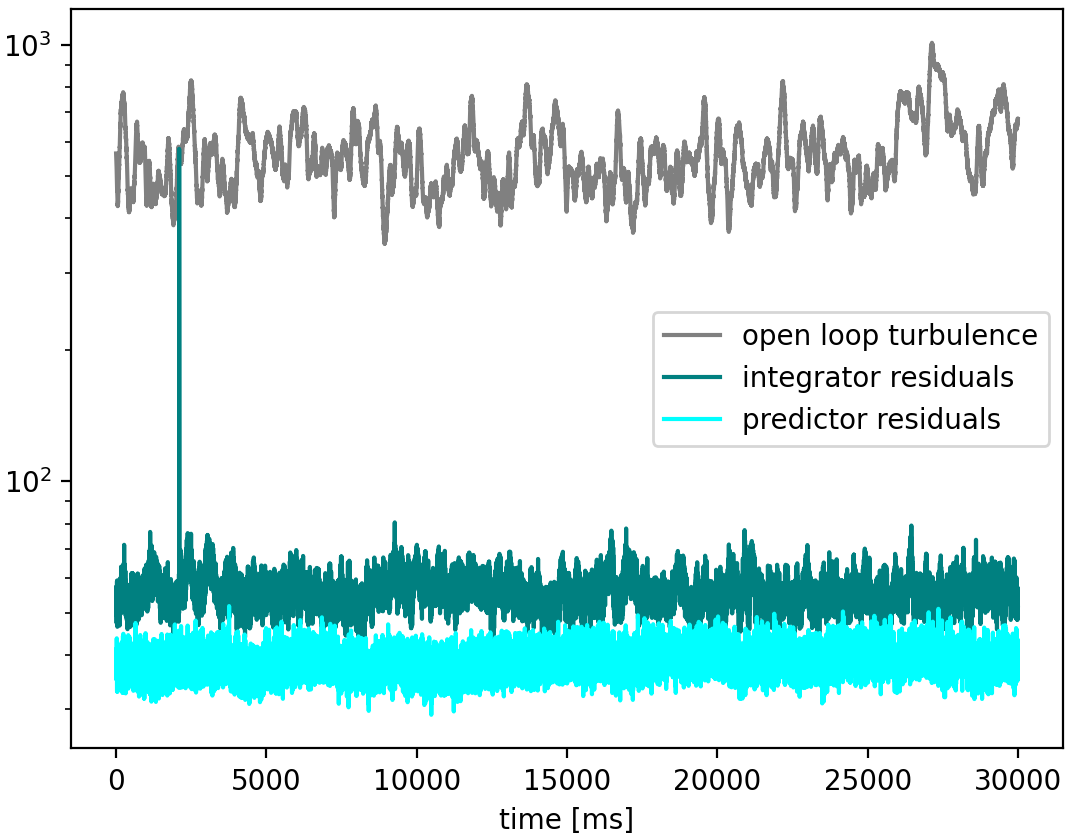}
   \end{tabular}
   \end{center}
   \caption[Predictor vs integrator performance on Keck-II AO bench.] 
%>>>> use \label inside caption to get Fig. number with \ref{}
   { \label{fig:on_bench_demo} 
Performance of the integrator and predictor as compared to the full state of injected turbulence on the Keck-II daytime bench. We show the RMS of the wavefront residual error as measured by the SHWFS with iterations of the controller. Left: Seven-layer atmospheric profile from Ref.  \citenum{KAON303}. On this high turbulence profile the gap between the improvement of the predictor over the integrator is more noticeable. Right: For this single profile with one simulated wind layer, there is still notable performance improvement offered by the predictor. }
   \end{figure}

We compare the root mean square (RMS) performance of the integrator with a high order gain of 0.5 and the predictor with a mixing factor and $\alpha$ of 1, which is only possible in simulation or in optimal conditions with lab data. Both the predictor and integrator were run at 1kHz. The data are compared in a collapsed time series (i.e., 0 is the start of each run for all three populations, but they are not the same 0) to show the comparable evolution in time of each controller and the open loop turbulence. The turbulence is limited to two minutes by the DM control software, and loops at the two minute mark, which creates a spike in wavefront residuals, as neither integrator nor predictor expect a fully disconnected scene of turbulence. We find promising bench results on both turbulence profiles, as shown in Figure \ref{fig:on_bench_demo}. The predictor consistently out-performs the integrator in this lab verification experiment by a factor of 1.7 for the higher turbulence case. 

%\section{ENGINEERING DATA WITH KECK-II AND NIRC2}

%\subsection{Observing Data}

%\begin{table}[ht]
%\caption{Margins and print area specifications.} 
%\label{tab:Paper Margins}
%\begin{center}       
%\begin{tabular}{|l|l|l|l|l|l|l|l|l|} 
%\hline
%\rule[-1ex]{0pt}{3.5ex}  Target & Mag (V) & Exposure Time & integrator gain & mixing factor & training data & history vector & $\alpha$ & use \\
%\hline
%\rule[-1ex]{0pt}{3.5ex}   & 2.54 cm & 1.0 in. & & & & & &  \\
%\hline

%\hline 
%\end{tabular}
%\end{center}
%\end{table}

%\subsection{Seeing and wind conditions on the night of 12/03/2025}

% # FIXME really just comment on the seeing (if mass/dimm was even working lol) and see if we can pull a ground layer out -- would be useful to comment on how this will be neat compared to faster wind layers. 

\section{On-sky results with Keck-II AO and NIRC2}

\subsection{Initial on-sky performance verification}

We performed initial verification in a half hour of observing on the target BS 8088, V magnitude = 6.26,  on the night of 10/5/2025, where we confirmed that we could successfully switch to the predictive control mode and close AO loops while maintaining nominal performance. In this initial test the $\alpha$ parameter was set to 2600, which results in a predictive filter that closely resembles an integrator. This was meant only as a first operational test. 

\begin{figure} [ht]
   \begin{center}
   \begin{tabular}{c} %% tabular useful for creating an array of images 
   \includegraphics[height=6cm]{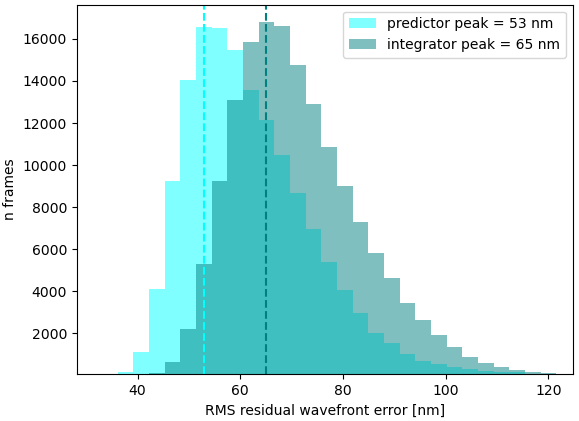}
   \includegraphics[height=6cm]{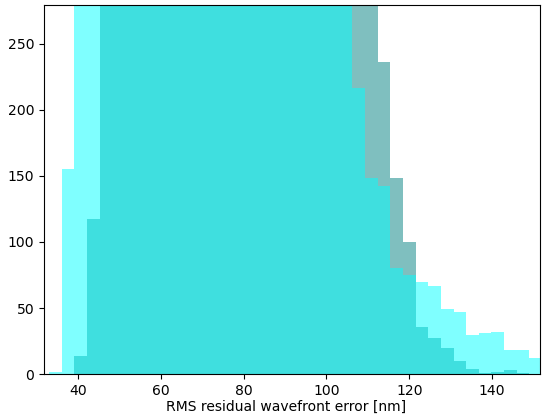}
   \end{tabular}
   \end{center}
   \caption[Predictor vs integrator performance during first on-sky test: residual wavefront on the SHWFS.] 
%>>>> use \label inside caption to get Fig. number with \ref{}
   { \label{fig:on_sky_demo_wfe_1} 
Histogram of RMS wavefront residuals during the initial predictor on/off experiment. Left: The mode of the two datasets are offset by 19$\%$, with the predictor outperforming the integrator by the measure of the SHWFS wavefront residuals. Right: A zoomed in view of the left figure; in this experiment we see a high noise tail from the prediction population, likely due to lower regularization parameter $\alpha$ and lower mixing factor making the correction less stable that optimal filter hyper-parameters.}
   \end{figure}

We performed our first quantitative test in an hour of observing on the target HD 189337, V magnitude = 6.5, the night of 10/19/2025, where we briefly explored possible filter parameters by manually scanning through filters, and ran consecutive predictor on/off experiments (predictor interleaved with integrator data) for a direct comparison between the predictor and integrator. For this experiment we used a high order integrator gain of 0.5, a mixing factor of 0.5, a low $\alpha$=250, a history vector (\textbf{n}) of 10 frames, and 60000 frames of training data (\textbf{l}), with both the integrator and predictor running at 1 kHz. Figure \ref{fig:on_sky_demo_wfe_1} shows these results; in $\sim$20 minutes of data we see a shift of 19$\%$ in the peak of the RMS wavefront error residuals sensed by the SHWFS towards a lower value. However, with these initial filter hyper-parameters we see a high noise tail. This is likely due to the higher mixing factor and a low regularization parameter $\alpha$, which creates a correction that is more aggressive and therefore more sensitive to data quality, as compared to optimal parameters from the second night, as discussed in Section \ref{sec:bench}. 

\begin{figure} [ht]
   \begin{center}
   \begin{tabular}{c} %% tabular useful for creating an array of images 
   \includegraphics[height=6cm]{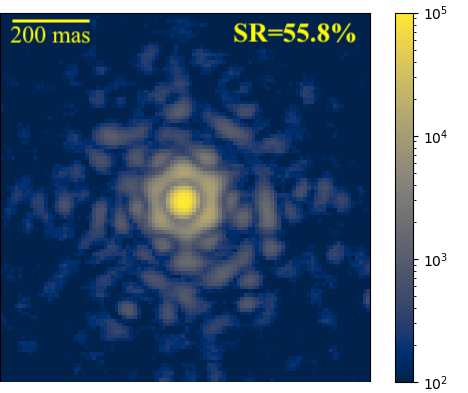}
      \includegraphics[height=6cm]{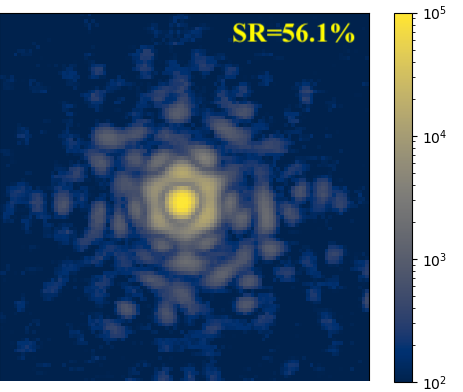}
   \end{tabular}
   \end{center}
   \caption[Predictor vs integrator performance during first on-sky test: NRC2 images.] 
%>>>> use \label inside caption to get Fig. number with \ref{}
   { \label{fig:on_sky_demo_psf_1} 
NIRC2 Bracket $\gamma$ images of the target HD 189337. We show a median combination of 5 0.2 ms frames. Left: Images while the integrator was running. Right: Images while the predictor was running. The performance in SR is comparable between the two median combined images.}
   \end{figure}

Along with measuring real time wavefront residuals, we include NIRC2 imaging through the Bracket $\gamma$ filter at 2190 nm. Figure \ref{fig:on_sky_demo_psf_1} shows the Strehl ratios (SR) on the median combined data with and without the predictor running, which show comparable performance around 56$\%$ SR. Data were minimally reduced, with bad pixel masking and dark subtraction. Strehl ratios are calculated using a modified version of method 7 from Ref.   \citenum{strehl}. 

%that obscures some of these performance gains. As discussed in Ref.   \cite{Fowler2024}, we expect that optimal hyper-parameters for PWFC will vary based on atmospheric parameters, and therefore we expect that better filter parameters will deliver an even greater performance boost.  

\subsection{Optimizing predictive filter hyper-parameters on-sky}
\label{sec:hyper-parameters}

On the night of 12/03/2025, we conducted a half night of Keck observing to conduct parameter scan experiments to quantify predictor performance and explore optimal hyper-parameters for the predictor on-sky. For these experiments we used a high order gain of 0.5 and ran the AO system at 1kHz. We used three targets, BS 8936, V magnitude = 6.64, BS 327, V magnitude $\sim$ 5.8, and BS 1191, V magnitude = 5.77. These targets were chosen to be bright natural guide stars that would experience low airmass during the duration of the observations. We designed the night with 4 experiments in mind: 
\begin{enumerate}
    \item{Conduct predictor on/off experiments for a quantitative comparison between integrator and predictor performance.}
    \item{Scan through mixing factors, to gauge what weighted combination of integrator and predictor performed best on-sky.}
    \item{Scan through regularization parameter $\alpha$, to gauge how aggressive the predictive filter should be for optimal performance.}
    \item{Scan through history vector and training data length, to gauge how much previous information produced the best predictive filter.}
\end{enumerate}

 In our mixing factor experiment (2), we found that any mixing factor higher than 0.4 resulted in an unstable loop, which was prone to high noise or loosing closed-loop performance; similarly, mixing factors less than 0.4 were not notably different than the integrator population. Varying $\alpha$ had a similar effect: any lower than an $\alpha$ of 300 caused the loop to go unstable (or be highly prone to going unstable), and higher values had the effect of washing out the predictor improvement, e.g., moving it more towards a integrator correction. For these experiments where the correction was unstable we did not collect consistent data, as intentionally collecting minutes of data on an unstable correction risks building up large DM shapes that impact quality the rest of night. %These results are shown breifly as a distinct population in Figure \ref{fig:on_sky_demo_wfe_2}. 
For this reason we focus the rest of this analysis on experiments (1) and (4). 

We compare the performance of various filter hyper-parameters by comparing the residual wavefront error sensed by the SHWFS. We take the RMS of the residual at each frame, and report that value as wavefront error. We note that on-sky the SHWFS is not sensitive to the entire wavefront error term (estimated to be an RMS error of 225 nm  \cite{KAON1322}). The SHWFS is blind to fitting error (error due to limited deformable mirror degrees of freedom) due to its sampling to a matching number of subapertures. Similarly, SHWFS signal should be noisier as measurement error (error due to low SNR on the SHWFS) increases, but may not reflect the exact predicted error term. Finally, the SHWFS has been shown to be insensitive to segment co-phasing errors in the primary mirror, and will not sense non-common path aberrations (NCPA) that occur in a path other than the SHWFS path. 

\begin{figure} [ht]
   \begin{center}
   \begin{tabular}{c} %% tabular useful for creating an array of images 
   \includegraphics[height=6cm]{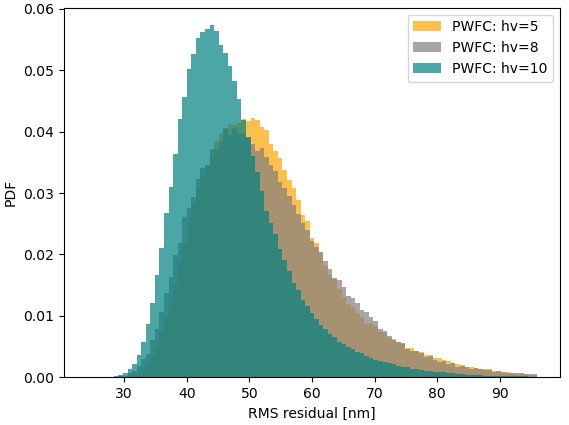}
      \includegraphics[height=6cm]{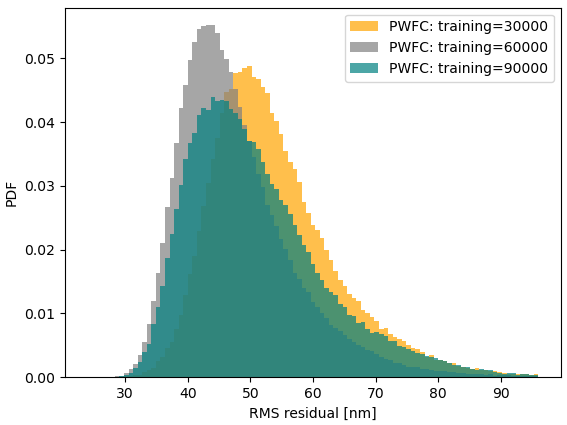}
   \end{tabular}
   \end{center}
   \caption[On-sky hyper-parameter optimzation for predictive control.] 
%>>>> use \label inside caption to get Fig. number with \ref{}
   { \label{fig:parameter_scan_experiments_data} 
Selecting optimal history vector (hv) and training data lengths for the predictive control filter. Using SHWFS wavefront sensor residuals during predictor periods running with different filter hyper-parameters, a clear set of optimal hyper-parameters emerge. Left: Data are sorted by history vector (number of frames used by the controller at every real time iteration) length. The longest history vector of 10 frames is preferred. Right: Data are sorted by training data (number of frames used to train the filter matrix) length. The median training value of 60000 frames is preferred, as more training data introduces noise into the controller. This difference speaks to how real noise in data impacts optimal training, but not history vector length. For both sets of histograms, the training data and history vector respectively vary while the other is held constant. Data on all three targets (BS 8936, BS 327, and BS 1191) are included. }
   \end{figure}

Figure \ref{fig:parameter_scan_experiments_data} shows the results from experiment (4), exploring how much data are used to train the predictive filter and how much data are used in real time to calculate the correction for the next step. We try history vector lengths of 5, 8, and 10, and see a clear preference for a history vector of 10 frames; with the current EOF implementation we are limited to 10 frames by the RTC. We try a training data length of 30000, 60000, 90000 frames; with a frame rate of 1kHz this amounts to a 0.5, 1, and 1.5 minutes of data. We see a clear preference for a training data length of 60000 frames. During this experiment more training data resulted in overfitting to noisy data, which diverges from similar experiments using simulated data, as explored in Ref. \cite{Fowler2024}. We use these filter parameters for the predictor on/off experiment (1), to compare integrator performance with optimal predictor performance. 

\subsection{Residual wavefront error and Strehl ratio as performance metrics}

We compare performance between the integrator with a high order gain of 0.5 and predictor performance, with AO running at 1kHz. We discuss three populations of predictor performance: (1) with a mixing factor of 0.4, a history vector length of 10, a training data length of 60000, and an $\alpha$ of 300, which were found to be the optimal filter hyper-parameters, as discussed in Section \ref{sec:hyper-parameters}, which we refer to as``static parameters'', (2) with a mixing factor of 0.4, varying history vector between 6-10, varying training data between 60000-1200000 frames, and an $\alpha$ of 300,  which we refer to as ``varying data length", and (3) a mixing factor of 0.4, a history vector length of 10, a training data length of 60000 frames, and a varying $\alpha$ between 1-1000, which we refer to as ``varying $\alpha$ or alpha''. 

\begin{figure} [ht]
   \begin{center}
   \begin{tabular}{c} %% tabular useful for creating an array of images 
   \includegraphics[height=6cm]{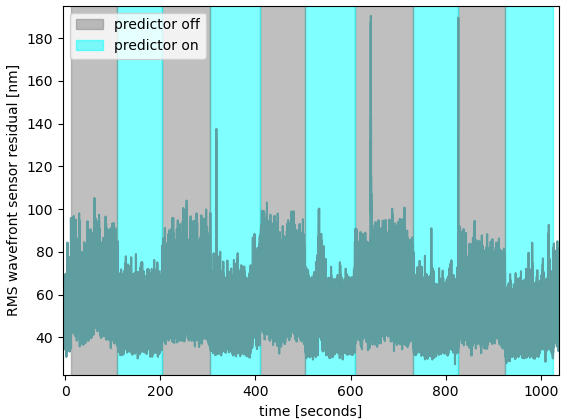}
      \includegraphics[height=6cm]{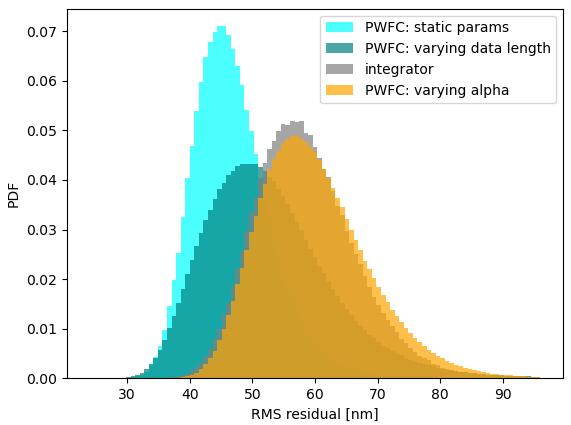}
   \end{tabular}
   \end{center}
   \caption[Predictor vs integrator performance during engineering half-night: residual wavefront on the SHWFS.] 
%>>>> use \label inside caption to get Fig. number with \ref{}
   { \label{fig:on_sky_demo_wfe_2} Data from the night of 12/03/25 on the target BS 1191. 
Left: RMS residual wavefront error from the SHWFS over time during the predictor on/off experiment. Cyan and grey regions indicate where the predictor and integrator were running respectively. Switching on the predictor consistently reduces the RMS wavefront error. Right: Histogram of the same predictor on and off population, as well as the performance of the predictor while exploring the filter hyper-parameter space. The optimal filter hyper-parameter population show a clear offset from the integrator population, (with a 20$\%$ improvement in RMS wavefront error) but even the populations with varying parameters show comparable to if not less wavefront error as compared to the integrator. }
   \end{figure}

We compare how the SHWFS wavefront residuals vary in time series during experiment (1) with fast switches between integrator and the predictor with static parameters. Figure \ref{fig:on_sky_demo_wfe_2} shows the notable improvement in wavefront error, as well as reduced noise while the predictor runs. Figure \ref{fig:on_sky_demo_wfe_2} also explores the same data, along with the predictors with varying data and varying alpha in the form of a histogram to examine where performance peaks. While both of the varying parameter populations do not perform as well as the static population, all of the predictor populations have a median performance in RMS wavefront error that is comparable to, if not less than the integrator. We note that extreme ($\alpha <$ 200) values can induce noise. These results are summarized in Table \ref{tab:wfe_numbers}. These results show that hyper-parameter optimization could be used to get the most out of predictor performance, but sub-optimal parameters are unlikely to degrade performance worse than a classic integrator. 

We note that the data from experiment (1) were collected while running with the same filter matrix during real on-sky turbulence. This implies that for the conditions on this night a single filter matrix was sufficient for up to $\sim$20 minutes, meaning that replacing a filter matrix every 5-10 minutes will likely be sufficient for standard on-sky operations. 

\begin{table}[ht]
\caption[Summary of controller performance.]{Numerical summary of controller performance according to SHWFS residuals.} 
\label{tab:wfe_numbers}
\begin{center}       
\begin{tabular}{|l|l|l|l|l|l|l|l|} 
\hline
\rule[-1ex]{0pt}{3.5ex}  Controller & Frames & m & n & l & $\alpha$ & Median & STD \\
\rule[-1ex]{0pt}{3.5ex}   &  &  & &  & & (nm) & (nm) \\
\hline
\hline
\rule[-1ex]{0pt}{3.5ex}  predictor & 470000 & 0.4 & 10 & 60000 & 300 & 46.1 & 6.3 \\
\hline
\rule[-1ex]{0pt}{3.5ex}  predictor & 4520000 & 0.4 & 6-10 & 30000-90000 & 300 & 51.4 & 10.2 \\
\hline
\rule[-1ex]{0pt}{3.5ex}  predictor & 3090000 & 0.4 & 10 & 60000 & 1-1000 & 59.0 & 62.3 \\
\hline
\rule[-1ex]{0pt}{3.5ex}  integrator & 484000 & N/A & N/A & N/A & N/A & 57.8 & 8.4 \\
\hline 
\end{tabular}
\end{center}
\end{table}

\begin{figure} [ht]
   \begin{center}
   \begin{tabular}{c} %% tabular useful for creating an array of images 
   \includegraphics[height=6cm]{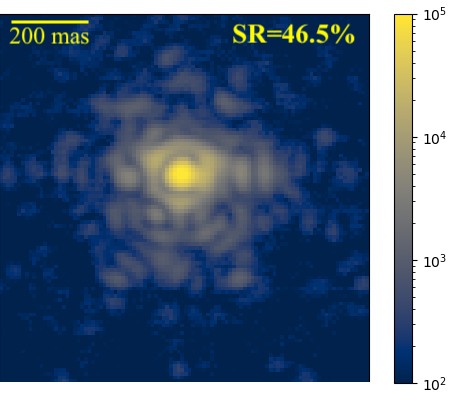}
      \includegraphics[height=6cm]{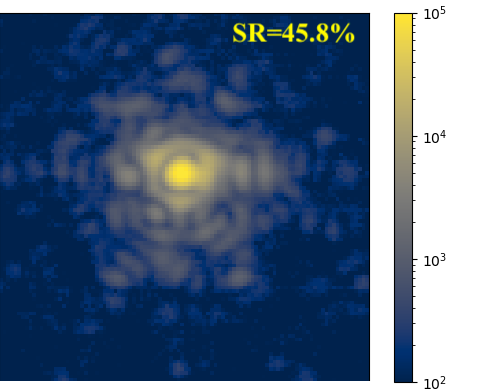}
   \end{tabular}
   \end{center}
   \caption[Predictor vs integrator performance during engineering half-night: NIRC2 images.] 
%>>>> use \label inside caption to get Fig. number with \ref{}
   { \label{fig:on_sky_demo_psf_2} 
 Keck/NIRC2 data through the Bracket $\gamma$ filter from the night of 12/03/2025. These data consist of a median combination of 40 0.1 second frames on the target BS 1191 while (left) a classic integrator was running and (right) a predictor with optimal filter hyper-parameters was running. }
   \end{figure}

\begin{figure} [ht]
   \begin{center}
   \begin{tabular}{c} %% tabular useful for creating an array of images 
   \includegraphics[height=6.3cm]{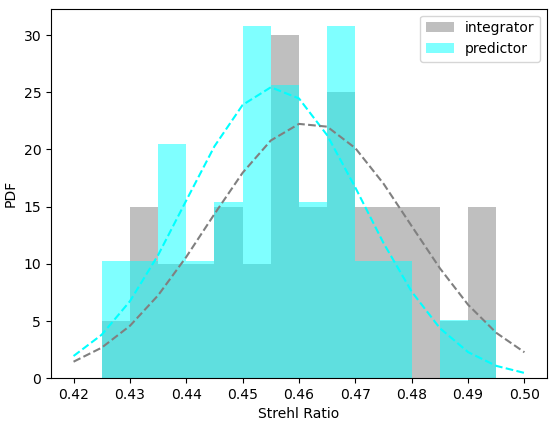}
   \end{tabular}
   \end{center}
   \caption[Predictor vs integrator performance during engineering half-night: SRs from imaging data.] 
%>>>> use \label inside caption to get Fig. number with \ref{}
   { \label{fig:on_sky_demo_image_metrics} 
Strehl ratios of the Keck/NIRC2 images taken during the predictor on/off experiment. Each NIRC2 frame is 0.1 seconds long on the target BS 1191. The integrator shows $<1\%$ improvement in SR over the predictor, which is less than the population standard deviation and likely negligible. %Right: Coronagraph-free contrast curves taken with Keck/NIRC2. By intentionally over-saturating the NIRC2 detector (0.5 seconds on an X target) we can roughly estimate comparative contrast between the two controllers. We estimate contrast using the \texttt{pyklip.klip.meas\_contrast} \cite{pyklip} function applied to a median combination of 5 0.5 s frames. The integrator and predictor show comparable contrast with this proxy method.    
}
   \end{figure}

We also present NIRC2 imaging data corresponding to predictor on/off periods shown in Figure \ref{fig:on_sky_demo_wfe_2}. Figure \ref{fig:on_sky_demo_psf_2} shows a median combination of Bracket $\gamma$ NIRC2 exposures, and an SR for each predictor on/off image, calculated with a modified version of method 7 from Ref.   \citenum{strehl}. The SRs between the integrator and predictor are comparable, around 46 $\%$. A histogram of the SR for each individual frame and a Gaussian fit to the data is shown in \ref{fig:on_sky_demo_image_metrics}. The peak in population for each individual frame is comparable between predictor and integrator, with a difference in population that is less half of a standard deviation. The difference as recorded by the SHWFS between the predictor on and off is $\sim$15 nm, which using an empirically adjusted Marechal approximation akin to Ref. \citenum{KAON1322} results in a difference of $\leq 1.5\%$ Strehl ratio, which would be hard to resolve in a seeing averaged night, especially when other factors are likely limiting Strehl ratio (e.g., 10$\%$ variation between the October and December night despite comparable wavefront sensor error measured from the SHWFS). 

\subsection{High contrast imaging performance metrics}

\begin{figure} [ht]
   \begin{center}
   \begin{tabular}{c} %% tabular useful for creating an array of images 
   \includegraphics[height=6cm]{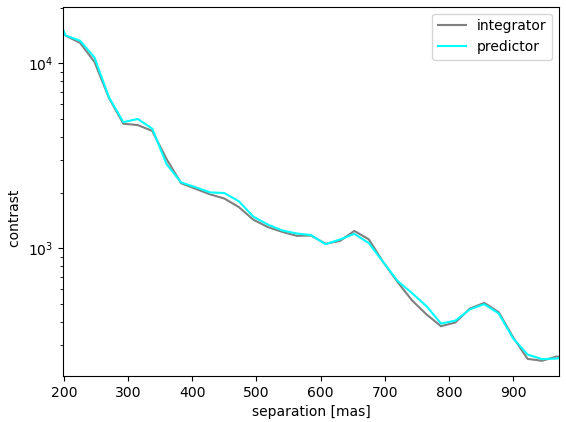}
      \includegraphics[height=6cm]{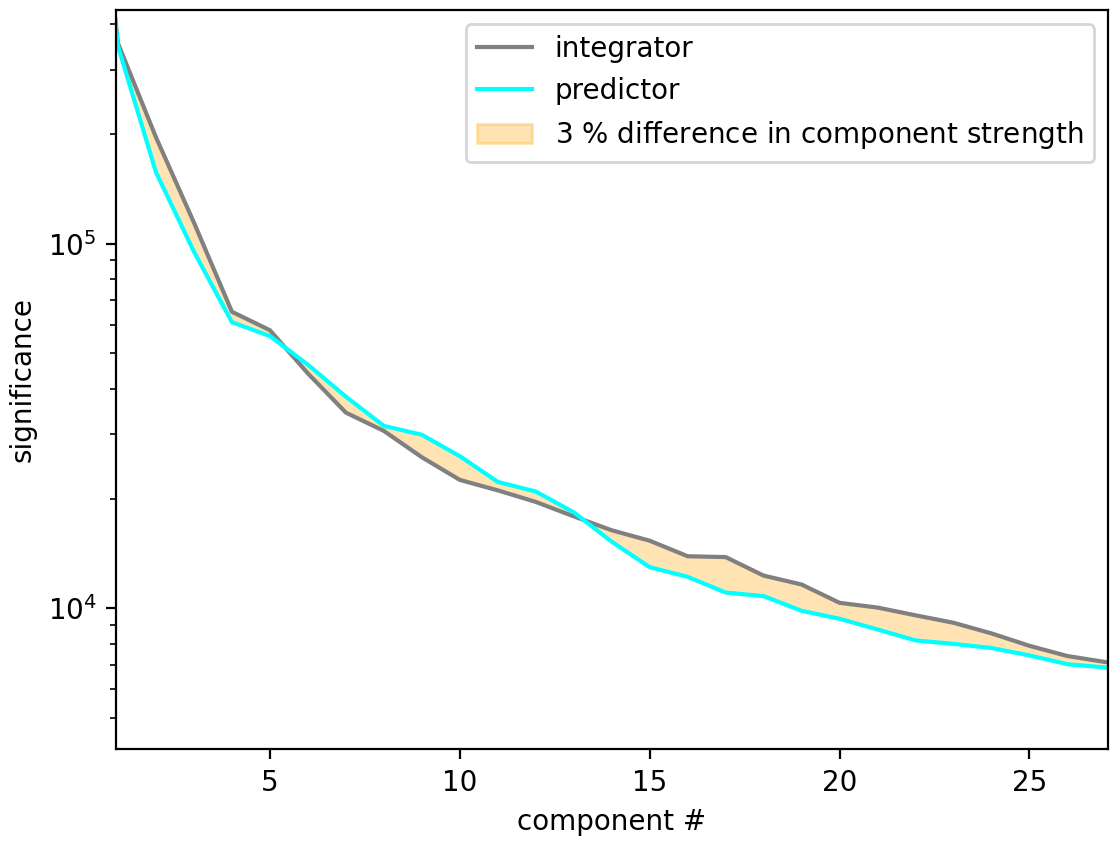}
   \end{tabular}
   \end{center}
   \caption[Predictor vs integrator performance during engineering half-night: high contrast imaging metrics.] 
%>>>> use \label inside caption to get Fig. number with \ref{}
   { \label{fig:on_sky_demo_hci_metrics} 
Left: Coronagraph-free contrast curves taken with Keck/NIRC2. By intentionally over-saturating the NIRC2 detector (0.5 seconds on the target BS 1191) we can roughly estimate comparative contrast between the two controllers. The integrator and predictor show comparable contrast with this proxy method. Right: Measuring PSF stability with the strength of singular values. We compare the singular values (i.e., principal components) of the two populations of images from the predictor and integrator. The data taken while the predictor was running shows 3$\%$ less strength than the integrator in singular values, implying that there is less noise and easier paths to data reduction for the predictor images. }
   \end{figure}

In the context of future high contrast imaging applications, we explore two metrics that (1) demonstrate coronagraph-free contrast (due to the time-consuming nature of coronagraphic observing in the context of a half night) and (2) speak to image stability for better PCA-style data reduction methods. 
In Figure \ref{fig:on_sky_demo_hci_metrics} we show a coronagraph-free contrast curve, by intentionally over-saturating the NIRC2 detector to explore detector response with separation from the bright star at the center. We estimate contrast on this over-saturated image using the \texttt{pyklip.klip.meas\_contrast}  \cite{pyklip} function applied to a median combination of 5 0.5 s frames. This contrast proxy also shows comparable performance between the predictor and the integrator. 

We also calculate the principal components of the 40 predictor and integrator images using \texttt{scipy.linalg.svd}  \cite{scipy} to explore PSF stability in the context of standard high contrast imaging reduction strategies. Here we see the predictor outperforming the integrator, with less power in most modes, with a net improvement of 3$\%$, also shown in  \ref{fig:on_sky_demo_hci_metrics}. High contrast imaging metrics show a subtle performance improvement as compared to a standard integrator, and future work will explore issues of Keck-II AO bench noise floor or model mismatch that may be limiting predictor performance. 

\section{HAKA + predictive wavefront control}

\begin{figure} [ht]
   \begin{center}
   \begin{tabular}{c} %% tabular useful for creating an array of images 
      \includegraphics[height=8cm]{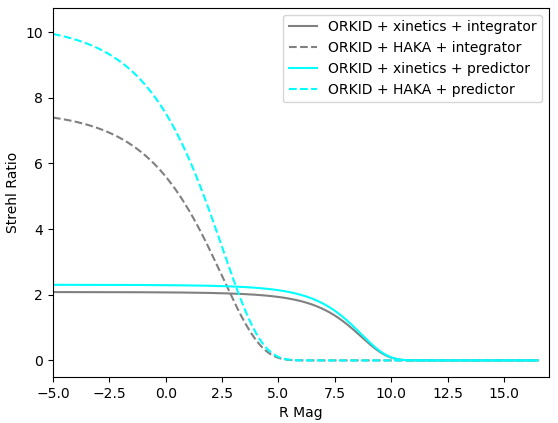}
   \end{tabular}
   \end{center}
   \caption[Project HAKA + ORKID performance using predictive control.] 
%>>>> use \label inside caption to get Fig. number with \ref{}
   { \label{fig:pwfc_orkid} 
In visible light, a few nm of wavefront has a larger impact. Using performance models of Keck/ORKID (the visible light Keck demonstrator instrument) we project the impact of the measured wavefront error improvement. Solid lines indicate the estimated performance using the xinetics deformable mirror and dashed lines show the predicted HAKA performance. }
   \end{figure}

At the time of this publication, Keck will have underdone an AO upgrade to use HAKA, a higher count actuator deformable mirror that will increase the current degrees of freedom in the Keck correction by up to a factor of 7. With fitting error dominating the pre-HAKA Keck AO error budget, after this upgrade temporal errors induced by control lag time will remain as one of the larger error terms. This work presents an opportunity to mitigate that error term with predictive control and ultimately deliver better performance for the next generation of Keck AO. Figure \ref{fig:pwfc_orkid} demonstrates this by simulating the impact on SR of the measured reduction in RMS wavefront error for HAKA feeding the visible light ORKID instrument. Using empirically tuned analytic models from Ref. \citenum{Fowler2026} we predict a performance improvement of 2 $\%$ SR on bright stars. Note that this only explores a consistent AO rate of 1kHz. 

However, HAKA's many actuators increase the size of the filter matrix, and threfore the training data needed for EOF. Work is still needed to ensure filter computation for a HAKA-sized filter matrix can happen in the few-minute timescale on which we expect atmosphere will stay consistent. Potential use of multiprocessing or GPU enabled filter computation, as well as strategic use of most significant singular values are potential solutions, but further work is needed to investigate potential performance loss from matrix inversion tools with less precision. 

\section{CONCLUSIONS AND FUTURE APPLICATIONS}

In conclusion, we present on-sky results of empirical orthogonal functions (EOF) as a linear data-driven predictive wavefront controller built into the Keck-II RTC. On-sky engineering tests of the predictive controller show a 20$\%$ performance improvement over a classic integrator according to wavefront residuals from the Shack-Hartmann Wavefront Sensor (SHWFS). Parameter optimization studies show that there is a clear improvement based on varying predictive filter hyper-parameters, but that within a reasonable regime, varying filter parameters does not degrade performance to notably worse than an integrator. Practically, this means that while hyper-parameter optimization does improve performance, parameters could be ``set and forget'' for ease of operation and this would not have an adverse effect on control. %We were not able to test simulated annealing as an optimal filter-finding strategy on-sky, but this method still shows promise for future start of night operations.

NIRC2 images through the Brackett $\gamma$=2190 nm filter show comparable performance between an integrator and predictor, both comparing SR and coronagraph-free contrast. We note that this is not similar to the up to 3x improvement in contrast in mid-spatial frequencies seen in Ref.  \citenum{vanKooten2022}, however the contrast we present here does not include use of the vortex coronagraph, which is highly sensitive to wavefront aberrations \cite{Guyon2006}. We also explore power in principal components, and find a modest improvement (on the order of 3$\%$ less area under the curve of component strength) from the predictor over the integrator. We expect that the Keck-II AO system at the time of these data was dominated by fitting error, as well as segment cophasing errors and non-common path aberrations. The impact of the latter two terms is demonstrated by the 10$\%$ difference in SRs between the October 19 night and the December 3 night, despite similar RMS wavefront error recorded by the SHWFS. 

With similar performance in wavefront error between the two nights (both seeing $\sim$ 50 nm RMS), it is also important to understand the full noise dynamics of the system and the true state of atmospheric turbulence. For example, predictive control performance gains may be limited by model mismatch (as discussed in Ref. \citenum{Fowler2023}), or we may be hitting the stability floor of the Keck-II AO system for a star of this magnitude. Future work will make a direct comparison to the noise floor of (1) the pseduo-open loop reconstruction of atmospheric turbulence and (2) the noise floor of the SHWFS to put these results in context. 

With HAKA on-sky at the time of this publication, and fitting error no longer the dominant term, temporal errors are predicted to be the dominant understood term (after the margin error term) in the Keck AO system, meaning that this improvement will be easier to estimate. Furthermore, visible light instruments are significantly more sensitive to wavefront errors, and empirically tuned analytic models of the visible light ORKID instrument  predict an improvement in SR from 7$\%$ (performance without prediction) to 9$\%$ (performance with prediction) for bright natural guide stars given the wavefront error improvement demonstrated here. 

Future work will explore the improvement of prediction at slower control frequencies. For faint guide stars, long exposures will be beneficial for reducing measurement error, but will incur greater temporal errors as the system lag time grows. A predictor could overcome this gap by predicting the state of the atmosphere after longer time lags, and will likely be the most beneficial for fainter guide stars. 

More on-sky demonstrations of novel controllers on large telescopes are necessary to bridge the gap between simulations, injected turbulence in a lab setting, and true on-sky turbulence and full telescope system complexity at a realistic outerscale. In the coming years SAXO+, the upgrade to the SAXO AO system for the Very Large Telescope/SPHERE will contribute to this goal by testing 6 predictors on the 8 meter VLT  \cite{ferreira2024}. But until that time Keck remains a one-of-a-kind testbed for the impact of predictive control on key science metrics at 8-10 meter class telescopes.

\appendix    %>>>> this command starts appendixes

\section{PREDICTIVE CONTROL BUSINESS UNIT UPDATE TO THE KECK-II REAL TIME CONTROLLER}
\label{sec:bu_update}

\begin{figure} [ht]
   \begin{center}
   \begin{tabular}{c} %% tabular useful for creating an array of images 
   \includegraphics[width=\textwidth]{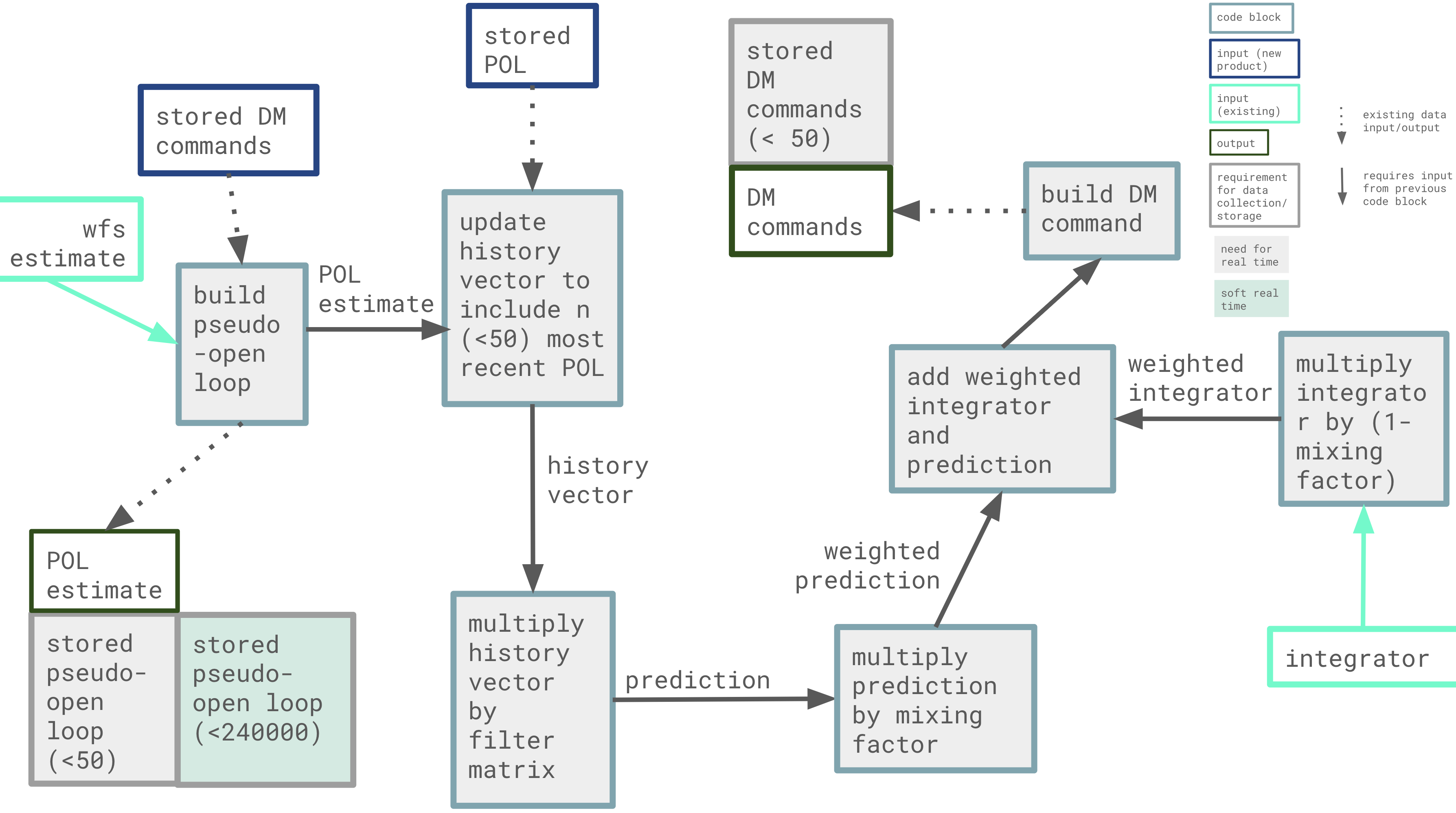}
   \end{tabular}
   \end{center}
   \caption[example] 
%>>>> use \label inside caption to get Fig. number with \ref{}
   { \label{fig:bu_block_diagram} 
Figure captions are used to describe the figure and help the reader understand it's significance.  The caption should be centered underneath the figure and set in 9-point font.  It is preferable for figures and tables to be placed at the top or bottom of the page. LaTeX tends to adhere to this standard.}
   \end{figure}

We commissioned a predictive control update to the Keck RTC. Figure \ref{fig:bu_block_diagram} shows how this update interacted with the existing RTC architecture. We specify new computations and data products required, as well as the required access to these products in soft and hard real time. We define hard real time as less than the AO frame rate, such that any computations can successfully run without slowing down the AO loop (expected to run at 1-2 kHz). We define soft real time as the 1-3 minute timescale on which we want to be able to successfully build and update a filter matrix. In particular, the predictive control business unit requires a new data type built into the Telemetry Recording Server (TRS) which contains the psuedo-open loop (POL) reconstruction for every real time iteration. 

%\section{DISCOVERY OF THE BS 1019 BINARY SYSTEM}

%BS 1019, known as a standard star from the Bright Star (BS) catalog, has never before been imaged with a 10 meter telescope. In attempting to use this as a calibration target, we discovered this is actually a binary system, with an estimated 40 mas separation. BS 1019 has previously been noted for its high acceleration in ..., but not directly confirmed as a binary system before these observations. 

\acknowledgments % equivalent to \section*{ACKNOWLEDGMENTS}       
 
The authors wish to recognize and acknowledge the very significant cultural role and reverence that the summit of Maunakea has always had within the Native Hawaiian community. We are most fortunate to have the opportunity to conduct observations from this mountain.

Some of the data presented herein were obtained at Keck Observatory, which is a private 501(c)3 non-profit organization operated as a scientific partnership among the California Institute of Technology, the University of California, and the National Aeronautics and Space Administration. The Observatory was made possible by the generous financial support of the W. M. Keck Foundation.

We acknowledge Heising-Simons Grant 2021-2373, National Science Foundation Grant 2008822, and National Science Foundation Grant A19-0912.  

% References
\bibliography{report} % bibliography data in report.bib
\bibliographystyle{spiebib} % makes bibtex use spiebib.bst

\end{document}